\documentclass{ws-procs975x65}

\def\beq{\begin{equation}}
\def\eeq{\end{equation}}

\begin{document}

\title{Radial excitations of non-static $J=0$ black holes in Einstein-Maxwell-Chern-Simons gravity}
\author{Jose Luis Bl\'azquez-Salcedo$^*$, Jutta Kunz}
\address{Department of Physics, University of Oldenburg,\\
Oldenburg, 26111, Germany\\
$^*$E-mail: jose.blazquez.salcedo@uni-oldenburg.com\\
}

\author{Francisco Navarro-L\'erida} 
\address{Departamento de F\'isica At\'omica, Molecular y Nuclear,\\ Universidad Complutense, Madrid,
28040, Spain\\
}

\author{Eugen Radu} 
\address{Departamento de F\'isica, Universidad de Aveiro, Aveiro,
3810-183, Portugal\\
}
%

\begin{abstract}
We study the generalization of the Kerr-Newmann black hole in 5D Einstein-Maxwell-Chern-Simons theory with free Chern-Simons coupling parameter. These black holes possess equal magnitude angular momenta and an event horizon of spherical topology. We focus on the extremal case with zero temperature. We find that, when the Chern-Simons coupling is greater than two times the supergravity case, new branches of black holes are found which violate uniqueness. In particular, a sequence of these black holes are non-static radially excited solutions with vanishing angular momentum. They approach the Reissner-Nordstr\"om solution as the excitation level increases. 
\end{abstract}


\bodymatter


\section{Introduction}

It is well known that in four space-time dimensions black holes satisfy the uniqueness theorem. The theorem states that asymptotically flat non-degenerate black holes in Einstein-Maxwell theory are uniquely characterized by the global charges such as the mass, the angular momentum and the electric charge. Rotating black holes with electric charge are described by the Kerr-Newman solution, which is known in closed form.

In higher dimensions this is no longer the case. The Schwarzschild, the Reissner-Nordstr\"om, and the Kerr metric have closed-form generalizations. But the rotating and electrically charged black hole in higher dimensional Einstein-Maxwell theory is not known analytically.

Nevertheless, other analytical solutions have been obtained in more general theories. One of these examples is found when a Chern-Simons term is added to the Einstein-Maxwell action in five dimensions \cite{CCLP05}. For a specific value of the Chern-Simons coupling (supergravity), the general rotating and electrically charged black holes are known analytically. A subset of these solutions are the BMPV black holes \cite{BMPV97}. These black holes have vanishing angular velocity, but they are not static since the total angular momentum does not vanish.

In this paper we are interested in the generalization of these EMCS black holes to general CS coupling constant. Since no general analytical solution is known, we will use numerical methods to generate the black holes and study their properties. 
In addition to the numerical methods, we can obtain some properties 
of extremal EMCS black holes if we study near-horizon solutions
in the entropy function formalism
\cite{SEN05}. In particular we will be interested in the solutions obtained when the CS coupling is greater than two times the supergravity value. We will see that for these values of the CS coupling, an interesting new family of solutions is found \cite{BS14,BS15}. 


Let us start presenting the theory and Ansatz.

\section{The Theory and the Ansatz}
\subsection{Theory}

Einstein-Maxwell-Chern-Simons theory in five dimensions has the following action:

\begin{equation} \label{EMCSac}
I= \frac{1}{16\pi G_5} \int d^5x\biggl[ 
\sqrt{-g} \, (R -
\frac{1}{4}F_{\mu \nu} F^{\mu \nu}
)
-
\frac{\lambda}{12\sqrt{3}}\,\varepsilon^{\mu\nu\alpha\beta\gamma}A_{\mu}F_{\nu\alpha}F_{\beta\gamma} 
 \biggr ],
\end{equation}
where $R$ is the curvature scalar and $A_\mu $ is the gauge potential  with field strength tensor $ F_{\mu \nu}
= \partial_\mu A_\nu -\partial_\nu A_\mu $. $\lambda$ is the CS coupling parameter
which in principle is free. Minimal 5-dimensional gauged supergravity is found for $\lambda=\lambda_{\rm SG}=1$. We choose the normalization so that $16\pi \, G_5=1$. Here we are interested in asymptotically flat space-times.  From this action we obtain the field equations:
\begin{equation}
\label{Einstein_equation}
G_{\mu\nu}=\frac{1}{2} F_{\mu\rho} {F_\nu}^\rho 
  - \frac{1}{8} g_{\mu \nu} F_{\rho \sigma} F^{\rho \sigma},\ \ \ \ \ \nabla_{\nu} F^{\mu\nu} + \frac{\lambda}{4\sqrt{3}}\varepsilon^{\mu\nu\alpha\beta\gamma}F_{\nu\alpha}F_{\beta\gamma}=0.
\end{equation}

\subsection{Ansatz}

We are interested in the 5-dimensional generalization of the Kerr-Newmann black holes, i.e., stationary black holes with spherical horizon topology and axial symmetry. Hence the space-time has the Killing vectors $\xi=\partial_t$, $\eta_{(1)}=\partial_{\varphi_1}$ and $\eta_{(2)}=\partial_{\varphi_2}$, where $t$ is a time-like coordinate, and $\varphi_1$, $\varphi_2$ are two angular coordinates related to the two axes of rotation. We restrict the black holes to have both angular momenta of equal magnitude,
$|J_{(1)}|=|J_{(2)}|=J$. This means that the symmetry is enhanced to cohomogeneity-1 configurations. All theses properties can be parametrized with the following Ansatz for the metric
\begin{eqnarray}
\label{metric}
&&ds^2 = -f(r) dt^2 + \frac{m(r)}{f(r)}(dr^2 + r^2 d\theta^2)  + \frac{n(r)}{f(r)}r^2 \sin^2\theta \left( d \varphi_1 -\frac{\omega(r)}{r}dt
\right)^2 \nonumber \\  && + \frac{n(r)}{f(r)}r^2 \cos^2\theta \left( d \varphi_2
  -\frac{\omega(r)}{r}dt \right)^2 + \frac{m(r)-n(r)}{f(r)}r^2 \sin^2\theta \cos^2\theta(d \varphi_1  -d \varphi_2)^2,
\end{eqnarray}
where $\theta \in [0,\pi/2]$, $\varphi_1 \in [0,2\pi]$ and $\varphi_2 \in [0,2\pi]$.
In addition, the Ansatz for the gauge potential is
\begin{equation}
A_\mu dx^\mu  = a_0(r) dt + a_{\varphi}(r) (\sin^2 \theta d\varphi_1+\cos^2 \theta d\varphi_2).
\end{equation}
The unknown metric and gauge potential functions depend only on the radial coordinate $r$, which we will assume to be a quasi-isotropic radial coordinate. 

The event horizon is located at $r=r_H$, where $f(r_H)=0$. And in particular extremal black holes are characterized by $r_H=0$, $f'(0)=0$.

\subsection{Charges and other properties}
Black holes can be characterized by several properties like their global charges. The existence of a time-like and angular Killing vectors allows us to use the the Komar formula to obtain the total mass $M$ and angular momentum $J$ of a configuration
\begin{equation}
M = -  \frac{3}{2} \int_{S_{\infty}^{3}} \alpha,  \ \ \ \ J_{(k)} =   \int_{S_{\infty}^{3}} \beta_{(k)} 
\ . \label{Kmass} \end{equation}
Where $\alpha_{\mu_1 \mu_2 \mu_3} \equiv \epsilon_{\mu_1 \mu_2 \mu_3
  \rho \sigma} \nabla^\rho \xi^\sigma$ and	
	%
	%
$\beta_{ (k) \mu_1 \mu_2 \mu_3} \equiv \epsilon_{\mu_1 \mu_2 \mu_3
  \rho \sigma} \nabla^\rho \eta_{(k)}^\sigma$, $k=1, 2$.
We can also calculate the electric charge $Q$ of the black hole,
\begin{equation}
Q= - \frac{1}{2} \int_{S_{\infty}^{3}} \tilde F 
\ , \label{charge} \end{equation}
where
${\tilde F}_{\mu_1 \mu_2 \mu_3} \equiv  
  \epsilon_{\mu_1 \mu_2 \mu_3 \rho \sigma} F^{\rho \sigma}$.

The event horizon rotates with angular velocity $\Omega_{\rm H}$, defined as
\begin{eqnarray}
\Omega_{\rm H} = \frac{\omega(r_{\rm H})}{r_{\rm H}}.
\end{eqnarray}

Other quantities of interest are related to the event horizon. For example, the area of the event horizon $A_{\rm H}$ and the horizon angular momentum $J_{{\rm H} (k)}$ are given by
\begin{equation}
A_{\rm H}=\int_{{\cal H}} \sqrt{|g^{(3)}|}=r_{\rm H}^{3} A(S^{3}) \lim_{r \to r_{\rm H}}
 \sqrt{\frac{m^{2} n}{f^{3}}}, \ \ \ \ \ J_{{\rm H} (k)} =   \int_{{\cal H}} \beta_{(k)} \label{hor_area} , \end{equation}
where ${\cal H}$ represents the surface of the horizon.
We define the
horizon angular momenta
$J_{{\rm H} (k)}$ by the Komar expression evaluated at the horizon. Note that equal-magnitude total angular momenta, $|J_{(k)}| =J$, implies equal horizon angular momenta,
$|J_{{\rm H} (k)}| =J_{\rm H}$.
Also note that the area is related to the entropy: $S = 4\pi A_{\rm H}$.

%

\section{Solutions Predicted by the Attractor Mechanism for $\lambda>2$}
Let us assume that the extremal black holes with event horizon of spherical topology have a near-horizon space-time with isometries given by $AdS_2 \times S^{D-2}$. An Ansatz incorporating these symmetries \cite{SEN05,BS14,BS15} is
\begin{eqnarray}
\label{metric_ansatz}
ds^2 &=& v_1(d\hat{r}^2/\hat{r}^2-\hat{r}^2dt^2) + v_2[4d\theta^2+\sin^22\theta(d\varphi_2-d\varphi_1)^2]
\\ 
\nonumber
&+&v_2\eta[d\varphi_1+d\varphi_2+\cos{2\theta}(d\varphi_2-d\varphi_1)-\alpha \hat{r} dt]^2.
\end{eqnarray}
Note that the horizon is at $\hat{r}=0$.
For the gauge potential we write
\begin{eqnarray}
\label{gv_ansatz}
A &=& -(\rho + p\alpha)\hat{r}dt + 2p(\sin^2\theta d\varphi_1 
+ \cos^2\theta d\varphi_2).
\end{eqnarray}
The Ansatz is described by the constant parameters $v_1$, $v_2$, $\eta$, $\alpha$, $\rho$ and $p$. These parameters  
satisfy some constraints which can be obtained from the field equations, or equivalently from the near-horizon formalism. The algebraic relations are
\begin{eqnarray}
\label{relgen}
&&v_2 = v_1, \nonumber\\
&&\eta v_1 = -\frac{4}{3}\frac{(\rho-p+p\alpha)(\rho+p+p\alpha)}{\alpha^2-1}, \nonumber\\
&&v_1 = \frac{2}{3}\frac{\alpha^4p^2-p^2+2\alpha^3\rho p-4\rho p \alpha + \alpha^2\rho^2 - 2\rho^2}{\alpha^2-1}, \nonumber
\\
&&3\alpha v_2^{5/2}\sqrt\eta(\rho+p\alpha) - 4\lambda\sqrt 3 p v_1 v_2 \rho - 4 \lambda \sqrt 3 p^2 v_1 v_2 \alpha  - 3p v_1^2\sqrt v_2 \sqrt \eta = 0,
\end{eqnarray}
and leave two undetermined parameters, which can be related to the angular momentum $J$ 
and the electric charge $Q$ of the extremal black hole. These two charges can be calculated using the associated Noether charges \cite{BS15,SW07}. 
The angular momentum $J$ is
\begin{eqnarray}
J 
= 64\pi^2\frac{v_2^{3/2}}{v_1}\sqrt \eta p(\rho+p\alpha) + 16\pi^2\frac{v_2^{5/2}}{v_1}\eta^{3/2}\alpha - \frac{256}{9}\sqrt 3 \pi^2p^3\lambda,
\end{eqnarray}
and the electric charge $Q$ is
\begin{eqnarray}
Q 
= -64\pi^2\frac{v_2^{3/2}}{v_1}\sqrt \eta (\rho+p\alpha) + \frac{128\pi^2\sqrt 3}{3}\lambda p^2.
\end{eqnarray}

We can calculate some horizon charges, such as the horizon angular momentum (from the Komar formula) and horizon area:
\begin{eqnarray}
J_{\rm H} 
=  16\pi^2\frac{v_2^{5/2}}{v_1}\eta^{3/2}\alpha, \ \ \ \ \ \ A_{\rm H} 
= 16\pi^2v_2^{3/2}\sqrt \eta.
\end{eqnarray}

\section{Global solutions in $\lambda>2$}

To obtain global solutions, we solve the EMCS equations numerically. We specify the usual boundary conditions for regular event horizon and asymptotic flatness. We use a numerical package implementing a collocation method for boundary-value
ordinary differential equations, with an adaptive
mesh selection procedure \cite{BS14,BS15}. 

Let us present results for $\lambda=5$ and fixed electric charge $|Q|=1$. The properties obtained for these parameters are generic for other values of $|Q|$ and $\lambda>2$.

\subsection{Branch structure and radially excited solutions}

\begin{figure}[h]
\begin{center}
\includegraphics[width=2in,angle=-90]{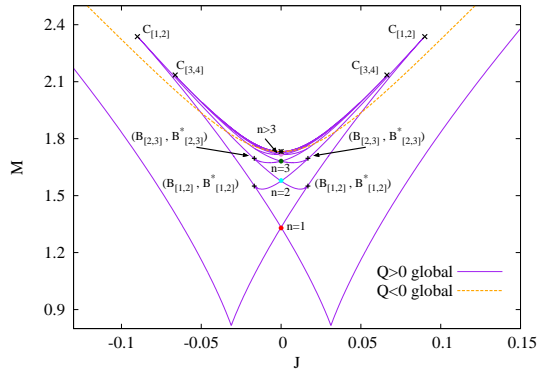}
\end{center}
\caption{Total mass vs total angular momentum for $\lambda=5$, $|Q|=1$ black holes. The branching points $(B, B^*)$ are marked with $+$, and the cusps $C$ with $\times$. The radially excited solutions are marked with $\bullet$, and the static solution with $\ast$.}
\label{fig1}
\end{figure}

The introduction of the CS term makes the theory no longer invariant under changes of sign of $Q$, for a fixed $\lambda$. For instance consider $Q = -1$. 
solutions. In Figure \ref{fig1} we can see that black holes with negative electric charge and $\lambda>2$ have always a finite horizon area, and
charged non-static solutions can be connected continuously with the charged static
solution by continuously decreasing the angular momentum. It can be demonstrated that the near-horizon solutions correspond one to one with the global solutions.

Black holes with $Q=1$ are completely different. A singular configuration with zero area is always found at some finite angular momentum. This corresponds to the peaks around $|J|=0.03$ in figure \ref{fig1}. Black holes with a larger $|J|$ have finite horizon area, and again the near-horizon solutions correspond one to one with the global solutions.
In Figure \ref{fig1} it can be seen that for values of $|J|$ between the singular solutions, a complicated branch structure is found. The structure contains cusps $C_{[i,j]}$, where the mass reaches a local maximum, but also of the area. The structure also contains branching points $B_{[i,j]},B^*_{[i,j]}$, where two different solutions with the same global charges coexist. The two solutions can be distinguished by their values of the horizon parameters (area, horizon angular momentum). This branch structure repeats an infinite number of times.
Note that in the branch structure, we find non-static $J=0$ solutions, and we have labeled them with an integer number $n$. This number corresponds to the number of nodes of the functions $\omega$ and $a_k$, and they can be understood as radially excited solutions. The mass of these solutions increases discretely towards the mass of the static black hole. Surprisingly, the near-horizon solution of this family of solutions is always the same.

\subsection{Domain of existence and non-uniqueness}

\begin{figure}[h]
\begin{center}
\includegraphics[width=2in,angle=-90]{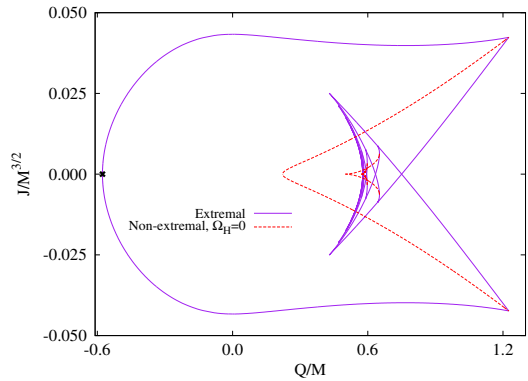}
\end{center}
\caption{Total angular momentum vs electric charge, both of them scaled to the mass. In red are the extremal solutions, which form the boundary of the domain of existence, except for the branch structure containing the radially excited solutions, which is found deep inside the boundary. In blue are the families of non-extremal solutions with $\Omega_H=0$. The intersection between the blue and red lines means non-uniqueness.}
\label{fig2}
\end{figure}


In figure \ref{fig2} we present the domain of existence of the black holes we have considered, representing $Q$ and $J$ scaled to the mass. Note that for positive $Q$, we have the branch structure of extremal solutions containing the radially excited black holes. These extremal black holes do not form the boundary of the domain of existence. In fact because of this, we can easily see in figure \ref{fig2} that there are non-extremal black holes with the same global quantities as these extremal solutions. In particular, we show solutions with $\Omega_H=0$, intersecting in multiple points the extremal solutions.

Hence, we can find two types of non-uniqueness. In the previous section we have seen non-uniqueness between extremal solutions ($B$ and $B^*$), at the branching points of figure \ref{fig1}). Now we can see that there is also non-uniqueness between extremal and non-extremal solutions. 

In addition we find that one near-horizon solution can correspond to only one global solution, multiple (even infinite) global solutions, or no global solution at all.

\section*{Acknowledgments}

We gratefully acknowledge support by the DFG Research Training Group 1620 “Models of Gravity” and by the Spanish MINECO, research project FIS2011-28013. The work of E.R. is supported by the
FCT-IF programme and the CIDMA strategic project UID/MAT/04106/2013.

\end{document}